\documentclass[twocolumn,showpacs,preprintnumbers,amsmath,amssymb,prb]{revtex4}
\usepackage{graphicx}% Include figure files
\usepackage{dcolumn}% Align table columns on decimal point
\usepackage{bm}% bold math

\begin{document}

\title{Effective permittivity of mixtures of anisotropic particles}

\author{M. Ya. Sushko}
\affiliation {Department of Theoretical Physics, Mechnikov
National University, \\
2 Dvoryanska St., Odessa 65026, Ukraine}

\email{mrs@onu.edu.ua}

\date{\today}

\begin{abstract}
We use a new approach to derive dielectric mixing rules for
macroscopically homogeneous and isotropic multicomponent mixtures
of anisotropic inhomogeneous dielectric particles. Two factors of
anisotropy are taken into account, the shape of the particles and
anisotropy of the dielectric parameters of the particles'
substances. Our approach is based upon the notion of macroscopic
compact groups of particles and the procedure of averaging of the
fields over volumes much greater than the typical scales of these
groups. It enables us to effectively sum up the contributions from
multiple interparticle reemission and short-range correlation
effects, represented by all terms in the infinite iterative series
for the electric field strength and induction. The expression for
the effective permittivity can be given the form of the
Lorentz-Lorenz type, which allows us to determine the effective
polarizabilities of the particles in the mixture. These
polarizabilities are found as integrals over the regions occupied
by the particles and taken of explicit functions of the principal
components of the permittivity tensors of the particles'
substances and the permittivity of the host medium. The case of a
mixture of particles of the ellipsoidal shape is considered in
detail to exemplify the use of general formulas. As another
example, Bruggeman-type formulas are derived under pertinent model
assumptions. The ranges of validity of the results obtained are
discussed as well.
\end{abstract}


\maketitle

\section{INTRODUCTION}
The study of effective permittivity of heterogeneous systems holds
an important position in various fields of physics and technics,
for its results find wide-spread applications in composite
material engineering, biochemical technology, medical diagnostics,
etc. In theoretical research, the simplest and much discussed
model considers a heterogeneous system as a mixture of fine
particles of the disperse phase embedded into a continuous host
medium. The development of it began with the case of dilute
mixtures of small spherical inclusions more than century ago
\cite{bib:MGarnett1904} and has resulted, in particular, in the
classical Maxwell-Garnett (MG) mixing rule and its various
modifications (see Ref.~\onlinecite{bib:Brosseau2006}  for an
analysis of relevant physical concepts and ideas from a historical
perspective, and Refs.~\onlinecite{bib:Bohren1983,
bib:Sihvola1999, bib:Tsang2001} for a review of major results). It
has been shown so far that: (1) The MG formula can incorporate
multiple scattering effects (see, for instance, analytical results
{\cite{bib:Claro1991, bib:Fu1993}} obtained in the quasistatic
limit within a mean-field approximation). (2) For certain
configurations of the disperse particles, it remains accurate for
high-concentrated mixtures, {\cite{bib:Abeles1976,
bib:Spanoudaki2001, bib:Wu2001, bib:Robinson2002, bib:Mallet2005}}
in which strong electromagnetic interaction is significant and for
which another -- the Bruggeman mixing rule
{\cite{bib:Bruggeman1935}} -- is often believed to be superior to
the former.

The case of mixtures of anisotropic particles remains
little-investigated (see reviews, {\cite{bib:Sihvola1999,
bib:Tsang2001}} after whose appearance the state of the art has
not changed much). As far as we know, the existing attempts at
taking the particles' anisotropy into account usually reduce to or
heavily rely on different kinds of one-particle approximations,
including their combinations (see, for instance,
Refs.~\onlinecite{bib:Levy1997, bib:Jilha2007} and references
therein). A typical example of such approximations is the use of
the one-particle polarizability, describing the response of a
solitary particle to a uniform electric field, instead of the
effective polarizability of the particle in the mixture. It is
evident that such an approach is tolerable only for diluted gases
of anisotropic particles. In sufficiently concentrated mixtures,
both multiple polarization effects  and many-particle correlations
in positions and orientations of the particles come into play. As
a result, finding the effective polarizability becomes a
many-particle problem, which is equivalent to the original problem
of finding the effective permittivity of the mixture.
Correspondingly, neither the effective polarizability as a
function of the dielectric and geometric parameters of disperse
particle, nor the interrelation of the two factors of anisotropy
of the particles -- nonsphericity of the shape and anisotropy of
the substance -- can be determined consistently within a
one-particle approximation. Yet we are unaware of any practically
important attempts at approaching these problems using the methods
of multiple-scattering theory (in contrast to the case of
spherical inclusions, for which see review, {\cite{bib:Tsang2001}}
key works, {\cite{bib:Lax1952, bib:Lamb1980, bib:Tsang1980,
bib:Davis1985}} and Refs.~{\onlinecite{bib:Mallet2005,
bib:Kuzmin2005}}). This fact is readily explained by the lack of
knowledge of an infinite set of the correlation functions for
concentrated systems of anisotropic particle. Even if such
information were available, practical calculations would be
extremely difficult and would probably be limited to estimations
of several corrections to the Born approximation.

Recently, \cite{bib:Sushko2007} we proposed a new approach to
analysis of the long-wavelength value of the effective
permittivity of finely dispersed mixtures. The idea was to avoid
excessive theoretical refining on polarization and correlation
processes that occur within the system on particle-size and
interparticle-distance scales by averaging their contributions out
over macroscopic regions reproducing the properties of the entire
system. The appropriate procedure is based upon the notion of
macroscopic compact groups of particles and the averaging
\cite{bib:LandauV8} of fields over volumes much greater than the
typical scales of these groups. By applying it, we carried out
\cite{bib:Sushko2007} a rigorous analysis of the effective
permittivity of a concentrated mixture of spherically symmetric
dielectric balls with piecewise-continuous radial permittivity
profile. Later, \cite{bib:Sushko2009} the method was applied to
systems comprising nonspherical inclusions with scalar
permittivity. It was also shown that both the MG and Bruggeman
mixing rules can be reconstructed with it.

In the present report, approach \cite{bib:Sushko2007} is developed
for macroscopically homogeneous and isotopic mixtures of
anisotropic dielectric particles whose dielectric properties are
described by permittivity tensors and which are embedded in a host
medium with constant scalar permittivity; the particles are
assumed to be hard, measurable, and, in general, inhomogeneous. It
is shown that the averaged contributions from all-order reemission
and short-range correlation effects within such a system can be
effectively summed up. As a result, the effective static
permittivity of the system is obtained as an explicit function of
the parameters of the model. It can be given the form of the
Lorentz-Lorenz type, the polarization properties of the particles
being characterized by their effective polarizabilities in the
mixture. The latter are found via the geometric and dielectric
parameters of the particles and the host medium; as an example, a
mixture of particles of the ellipsoidal shape is considered in
detail. Finally, Bruggeman-type formulas are shown to follow from
the general formulas under special choices of the effective
medium, and the ranges of validity of the results obtained are
discussed.

\section{\label{sec:basic} BASIC RELATIONS FOR ELECTRIC FIELD
AND INDUCTION}

\begin{figure}
\includegraphics[width=6cm]{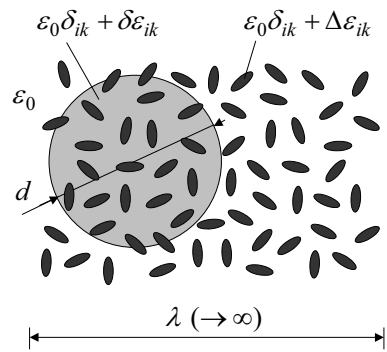}
\caption{\label{fig:f1}Homogenization of a mixture in terms of
compact groups, viewed as macroscopic many-particle regions with
linear sizes $d$ much less than the wavelength $\lambda$ of
probing radiation.}
\end{figure}

To begin with, we consider the problem on propagation of
electromagnetic waves in a finely dispersed mixture with local
permittivity $\varepsilon _{ik} ({\rm {\bf r}}) = \varepsilon _{0}
\delta _{ik} + \delta \varepsilon _{ik} ({\rm {\bf r}})$. Here,
$\varepsilon _{0} $ is the permittivity of the host medium and
$\delta \varepsilon _{ik} ({\rm {\bf r}})$ is the contribution
caused by compact groups of the disperse anisotropic particles. By
a compact group we understand any macroscopic region within which
all interparticle distances $ \vert {\rm {\bf r}}_{\lambda} - {\rm
{\bf r}}_{\mu} \vert$ are small as compared to the wavelength of
the probing wave in the mixture: $\sqrt {\varepsilon _{0}} k_{0}
\vert {\rm {\bf r}}_{\lambda} - {\rm {\bf r}}_{\mu} \vert < < 1$,
where $k_{0} $ is the wave vector of the wave in vacuum (see
Fig.~\ref{fig:f1}).
According to Ref.~\onlinecite{bib:Sushko2007}, we expect that it
is multiple reemission and short-range correlation effects within
such groups that form the effective permittivity of the mixture in
the static limit $\sqrt {\varepsilon _{0}}  k_{0} \to 0$, where
even quite large groups of disperse particles become compact. With
respect to an external static field, such groups can actually be
treated as point-like; fluctuations of macroscopically large
numbers of particles contained in them, as well as correlations
between the groups, can be ignored. Correspondingly, the deviation
$\delta \varepsilon _{ik} ({\rm {\bf r}})$ of the local
permittivity in the mixture from the permittivity of the host due
to the presence  at the point $\textbf{r}$ of a compact group of
$N$ identical disperse particles, occupying measurable regions
${\rm \Omega}_{\lambda}$ with individual volumes $v$, can be
modelled as
\begin{equation}
\label{eq:eq1} \delta \varepsilon _{ik} ({\rm {\bf r}}) =
{\sum\limits_{\lambda = 1}^{N} {\Delta \varepsilon _{ik} ({\rm
{\bf r}},{\rm \Omega}_{\lambda} )\,\Pi ({\rm {\bf r}},{\rm
\Omega}_{\lambda} )}} ,
\end{equation}
\noindent where $\Delta \varepsilon _{ik} ({\rm {\bf r}},{\rm
\Omega}_{\lambda} ) = \varepsilon _{ik} ({\rm {\bf r}},{\rm
\Omega}_{\lambda} ) - \varepsilon _{0} \delta _{ik} $ is the
deviation of the local permittivity in the mixture from the
permittivity of the host due to the presence of the $\lambda$th
particle at $\textbf{r}$, $\varepsilon _{ik} ({\rm {\bf r}},{\rm
\Omega}_{\lambda})$ stands for the permittivity tensor of the
substance of this particle, $\Pi ({\rm {\bf r}},{\rm
\Omega}_{\lambda} )$ is the characteristic function of the region
${\rm \Omega}_{\lambda}$ [$\Pi ({\rm {\bf r}},{\rm
\Omega}_{\lambda} ) = 1$ if $\textbf{r} \in {\rm
\Omega}_{\lambda}$ and $\Pi ({\rm {\bf r}},{\rm \Omega}_{\lambda}
) = 0$ otherwise], and $\delta _{ik}$ is the Kronecker delta. In
what follows, the tensors $\delta \varepsilon _{ik} ({\rm {\bf
r}})$, $\varepsilon _{ik} ({\rm {\bf r}},{\rm \Omega}_{\lambda}
)$, and $\Delta \varepsilon _{ik} ({\rm {\bf r}},{\rm
\Omega}_{\lambda} )$ will also be designated as $\hat {\mathop
{\delta \varepsilon} } $, $\hat {\mathop {\varepsilon} } $, and
$\hat {\mathop {\Delta \varepsilon} } $ respectively.

The equation for the $i$ component of the electric field of a wave
in the mixture can be written as \cite{bib:LandauV8}
\begin{equation}
\label{eq:eq2} \Delta E_{i} + k_{0}^{2} \varepsilon _{0} E_{i} -
\nabla _{i} \nabla _{k} E_{k} = - k_{0}^{2} \delta \varepsilon
_{ik} {\kern 1pt} E_{k};
\end{equation}
\noindent henceforth, the dummy suffix convention applies.
Eq.~(\ref{eq:eq2}) is equivalent to the integral equation
\begin{equation}
\label{eq:eq3} E_{i} ({\rm {\bf r}}) = E_{i}^{(0)} ({\rm {\bf r}})
- k_{0}^{2} {\int\limits_{V} {d {\rm {\bf {r}'}}\,T_{ij} ({\rm
{\bf r}},{\rm {\bf {r}'}})\delta \varepsilon _{jk} ({\rm {\bf
{r}'}})E_{k} ({\rm {\bf {r}'}})} },
\end{equation}
\noindent where  $E_{i}^{(0)} ({\rm {\bf r}}) = E_{i}^{0} \exp
\left( {i \sqrt {\varepsilon _{0}}  {\rm {\bf k}}_{0} \cdot {\rm
{\bf r}}} \right)$ is the $i$ component of the incident wave field
in the mixture, $E_{i}^{0}$ is the $i$ component of the field
amplitude ${\rm {\bf E}}_{0} $, and $T_{ij} (r) = {{ - \left(
{k_{0}^{2} \delta _{ij} + \nabla _{i} \nabla _{j}} \right)\exp
\left( {i k_{0} \sqrt {\varepsilon _{0}}}r \right)}
\mathord{\left/ {\vphantom {{ - \left( {k_{0}^{2} \delta _{ij} +
\nabla _{i} \nabla _{j}} \right)\exp \left( {{i} k_{0} \sqrt
{\varepsilon _{0} }} \right)} {4\pi k_{0}^{2} \varepsilon _{0}
r}}} \right. \kern-\nulldelimiterspace} {4\pi k_{0}^{2}
\varepsilon _{0} r}}$ are the components of the electromagnetic
field propagator [Green's tensor for Eq.~(\ref{eq:eq3})]; the
integral in Eq.~(\ref{eq:eq3}) is taken over the volume of the
mixture.

With the iterative procedure applied, the solution of
Eq.~(\ref{eq:eq3}) can be represented in the form
\begin{equation}
\label{eq:eq4} E_{i} ({\rm {\bf r}}) = E_{i}^{(0)} ({\rm {\bf r}})
+ {\sum\limits_{n = 1}^{\infty}  {E_{i}^{(n)} ({\rm {\bf r}})}} ,
\end{equation}
\begin{eqnarray}
E_{i}^{(n)} ({\rm {\bf r}}) = ( - k_{0}^{2} )^{n}{\int\limits_{V}
{d {\rm {\bf r}}_{1}}}  {\int\limits_{V} {d {\rm {\bf r}}_{2}}}
\ldots{\int\limits_{V} {d {\rm {\bf r}}_{n}}}  \,T_{ij} ({\rm {\bf
r}},{\rm {\bf r}}_{1} )\delta \varepsilon _{jk} ({\rm {\bf
r}}_{1})\nonumber \\
 \times  T_{kl} ({\rm {\bf r}}_{1} ,{\rm {\bf r}}_{2})
 \delta \varepsilon _{lm} ({\rm {\bf r}}_{2} )  \ldots  T_{rs}
({\rm {\bf r}}_{n - 1} ,{\rm {\bf r}}_{n} )\delta \varepsilon
_{st} ({\rm {\bf r}}_{n} )E_{t}^{(0)} ({\rm {\bf r}}_{n} ).
\nonumber\\
\label{eq:eq5}
\end{eqnarray}
\noindent The $i$ component of the electric induction vector at
the point ${\rm {\bf r}}$ in the mixture is given by
\begin{equation}
\label{eq:eq6} D_{i} ({\rm {\bf r}}) = {\left[ {\varepsilon _{0}
\delta _{ik} + \delta \varepsilon _{ik} ({\rm {\bf r}})}
\right]}\,E_{k} ({\rm {\bf r}}).
\end{equation}

Assuming that the mixture as a whole is macroscopically homogenous
and isotropic, we define its effective permittivity $\varepsilon
_{\rm eff}$ in the standard way, \cite{bib:LandauV8} as the
proportionality coefficient in the relation
\begin{equation}
\label{eq:eq7} \overline {{D}_{i}} = \varepsilon _{\rm eff} {\kern
1pt} \overline {{E}_{i}} ,
\end{equation}
\noindent where the bars indicate averaging by the rule $\overline
{{A}_{i}} = {\frac{{1}}{{V}}}{\int\limits_{V} {A_{i} ({\rm {\bf
r}}){\kern 1pt} d{\rm {\bf r}}}}$; the volume $V$ is implied to be
much greater than the volumes of compact groups. Following
Ref.~\onlinecite{bib:LandauV8}, we accept that for finely
dispersed mixtures, the values $\overline {{A}_{i}} $ are equal to
the corresponding statistical averages ${\left\langle {A_{i} ({\rm
{\bf r}})} \right\rangle} $ over the positions and orientations of
disperse particle.

Now, under the suggestions made, we can prove quantitatively the
following two statements: (1) In the limiting case $\sqrt
{\varepsilon _{0}} k_{0} \to 0$, the values ${\kern 1pt} \overline
{{E}_{i}} $ and $\overline {{D}_{i}} $ are determined only by the
multiple reemission and short-range correlation effects inside
compact groups. (2) The corresponding all-order contributions to
$\varepsilon _{\rm eff}$ can be singled out from the iterative
series by formal replacing the factors $k_{0}^{2} T_{ij} (r)$ in
the integrals for ${\kern 1pt} \overline {{E}_{i}} $ and
$\overline {{D}_{i}} $ with the expressions ${{\delta ({\rm {\bf
r}})\delta _{ij}} \mathord{\left/ {\vphantom {{\delta ({\rm {\bf
r}})\delta _{ij}} {3\varepsilon _{0}}} } \right.
\kern-\nulldelimiterspace} {3\varepsilon _{0}}} $, where $\delta
({\rm {\bf r}})$ is the Dirac delta function. This replacement
simply reflects the fact that within a macroscopic approach, the
specified contributions are formed by those ranges of the
integration variables where the electromagnetic field propagators
reveal a singular behavior.

Our proof uses the representation \cite{bib:LandauV4}
\begin{eqnarray}
{\mathop {\lim} \limits_{k_{0} \to 0}} k_{0}^{2} {\tilde T}_{ij}
({\rm {\bf r}}) = - {\frac{{1}}{{4\pi \varepsilon _{0}}} }\nabla
_{i} \nabla _{j} {\frac{{1}}{{r}}} \nonumber
 \\= {\frac{{1}}{{3\varepsilon _{0}}}
}\delta _{ij} \delta ({\rm {\bf r}}) + {\frac{{1}}{{4\pi
\varepsilon _{0} r^{3}}}}\left( {\delta _{ij} - 3e_{i} e_{j}}
\right) \label{eq:eq8}
\end{eqnarray}
\noindent for the propagator $T_{ij} ({\rm {\bf r}})$, which is
valid  (see Ref.~\onlinecite{bib:Sushko2004} for mathematical
details) on a set of scalar, compactly supported, and bounded
functions $\delta \varepsilon _{\alpha \beta} ({\rm {\bf r}})$ in
the sense that there holds the relation
\[
{\mathop {\lim} \limits_{k_{0} \to 0}} {\int\limits_{V} {d{\rm
{\bf r}}}} k_{0}^{2} T _{ij}({\bf{r}}) \delta \varepsilon _{\alpha
\beta} ({\rm {\bf r}}) = {\mathop {\lim} \limits_{k_{0} \to 0}}
{\int\limits_{V} {d{\rm {\bf r}}}} k_{0}^{2} {\tilde
T}_{ij}({\bf{r}}) \delta \varepsilon _{\alpha \beta} ({\rm {\bf
r}}).
\]
In Eq.~(\ref{eq:eq8}), $e_{i} $ stands for the $i$ component of
the unit vector ${\rm {\bf e}} = {{{\rm {\bf r}}} \mathord{\left/
{\vphantom {{{\rm {\bf r}}} {r}}} \right.
\kern-\nulldelimiterspace} {r}}$; for brevity, the first and the
second terms to the right of the equality sign will also be
denoted by $\hat {\tau} ^{(1)}$ and $\hat {\tau }^{(2)}$
respectively.

With the aid of (\ref{eq:eq8}), the contribution to the
statistical average ${\left\langle {E_{i} ({\rm {\bf r}})}
\right\rangle} $ from the $n$th iterative step can be represented
in the static limit $\sqrt {\varepsilon _{0}}  k_{0} \to 0$ as
\begin{equation}
\label{eq:eq9} {\left\langle {E_{i}^{(n)} ({\rm {\bf r}})}
\right\rangle}  = {\left\langle {E_{i}^{(n;1)} ({\rm {\bf r}})}
\right\rangle}  + {\left\langle {E_{i}^{(n;1,2)} ({\rm {\bf r}})}
\right\rangle}.
\end{equation}
The addend
\begin{equation}
\label{eq:eq10} {\left\langle {E_{i}^{(n;1)} ({\rm {\bf r}})}
\right\rangle}  = \left( { - {\frac{{1}}{{3\varepsilon _{0}}} }}
\right)^{n}{\left\langle {\delta \varepsilon _{ik} ({\rm {\bf
r}})\delta \varepsilon _{km} ({\rm {\bf r}}) \ldots \delta
\varepsilon _{rt} ({\rm {\bf r}})} \right\rangle }E_{t}^{0}
\end{equation}
represents the statistical average of a product of $n$ factors
$\hat {\mathop {\delta \varepsilon} }$ related to a single compact
group; it is obtained by replacing all factors $k_{0}^{2} \hat
{T}$ in (\ref{eq:eq5}) by the their most singular parts $\hat
{\tau }^{(1)}$. The addend ${\left\langle {E_{i}^{(n;1,2)} ({\rm
{\bf r}})} \right\rangle} $ is the sum of all integrals containing
at least one factor $\hat {\tau} ^{(2)}$ under the integral sign.
Each expression containing $m \leq n$ factors of the $\hat {\tau}
^{(2)}$ type results from taking $(n-m)$ integrals with
 $\hat {\tau }^{(1)}$ factors in their integrands;
it is therefore proportional to the expression
\begin{widetext}
\begin{eqnarray}
  {\int\limits_{V} {d {\rm {\bf r}}_{1}}}  {\int\limits_{V}
{d {\rm {\bf r}}_{2}}}  \ldots {\int\limits_{V} {d {\rm {\bf
r}}_{m}}} \,\hat {\tau }^{(2)}\left( { {\rm {\bf r}} - {\rm {\bf
r}}_{1} } \right)\hat {\tau} ^{(2)}\left( { {\rm {\bf r}}_{1} -
{\rm {\bf r}}_{2} } \right) \ldots  \hat {\tau} ^{(2)}\left( {
{\rm
{\bf r}}_{m - 1} - {\rm {\bf r}}_{m} } \right)\nonumber \\
 \times {\left\langle {\left( {\hat {\mathop {\delta \varepsilon} }
 ({\rm {\bf r}}_{1} )} \right)^{i_{1}}
\left( { \hat {\mathop {\delta \varepsilon} }  ({\rm {\bf r}}_{2}
)} \right)^{i_{2}} \ldots  \left( {\hat {\mathop {\delta
\varepsilon} }({\rm {\bf r}}_{m} )} \right)^{i_{m}}}
\right\rangle} {\rm {\bf E}}^{0}, \nonumber
 \end{eqnarray}
\end{widetext} \noindent where $i_{1} + i_{2} + ... + i_{m} =
n$. Since the spacial and orientational correlations between
macroscopic compact groups in a macroscopically homogeneous and
isotropic mixture are negligibly small, the many-point (involving
different compact groups) correlator in this expression can be
factorized as a product of one-point correlators, each of which is
related to a single compact group. The one-point correlators are
equal to linear combinations of expressions constructed of
Kronecker deltas and satisfying certain symmetry constraints on
permutations of their indices. In particular, the average
${\left\langle {\delta \varepsilon _{jk} ({\rm {\bf r}}_{1} )}
\right\rangle } = A\delta _{jk} $ is symmetric with respect to
permutations $j \leftrightarrow k$, the average ${\left\langle
{\delta \varepsilon _{jk} ({\rm {\bf r}}_{1} )\delta \varepsilon
_{lm} ({\rm {\bf r}}_{1} )} \right\rangle} = B\delta _{jk} \delta
_{lm} + C\left( {\delta _{jl} \delta _{km} + \delta _{jm} \delta
_{kl}}  \right)$ is symmetric with respect to permutations $j
\leftrightarrow k$, $l \leftrightarrow m$ of individual indices
and permutations $(j,k) \leftrightarrow (l,m)$ of their pairs,
etc. The coefficients $A$, $B$, $C, \ldots$ in these relations
depend on the physical parameters of the host and disperse
particles, but not on the coordinates of the relevant compact
group. As a consequence, the integrals in ${\left\langle
{E_{i}^{(n;1,2)} ({\rm {\bf r}})} \right\rangle} $ reduce to those
taken of factors $\hat {\tau} ^{(2)}$ alone. Taking into account
the explicit form of $\hat {\tau} ^{(2)}$, we see that these
integrals vanish after integration with respect to the angles.
Thus,
\begin{equation}
\label{eq:eq11} {\left\langle {E_{i}^{(n;1,2)} ({\rm {\bf r}})}
\right\rangle}  = 0.
\end{equation}

In view of the equality of the averages ${\left\langle {E_{i}
({\rm {\bf r}})} \right\rangle} $  and $\overline {{E}_{i}}$,
Eqs.~(\ref{eq:eq9})--(\ref{eq:eq11}) give
\begin{equation}
\label{eq:eq12} \overline {\mathop {E_{i}^{(n)} ({\rm {\bf r}})}}
= \left( { - {\frac{{1}}{{3\varepsilon _{0}}} }}
\right)^{n}\overline {\mathop {\delta \varepsilon _{ik} ({\rm {\bf
r}})\delta \varepsilon _{km} ({\rm {\bf r}})  \ldots  \delta
\varepsilon _{rt} ({\rm {\bf r}})}}E_{t}^{0} .
\end{equation}

Similar reasoning is used to analyze the averages ${\left\langle
{D_{i} ({\rm {\bf r}})} \right\rangle} $ and $\overline {{D}_{i}}
$. From Eqs.~(\ref{eq:eq4})--(\ref{eq:eq6}), we finally obtain:
\begin{widetext}
\begin{equation}
\label{eq:eq13} \overline {{E}_{i}} = \left[ \delta _{it} +
\sum\limits_{\sigma = 1}^{\infty} \left(  - \frac{1}{3\varepsilon
_{0}}  \right)^{\sigma} \overline {\mathop \delta \varepsilon
_{ik} ({\rm \bf r})\delta \varepsilon _{km} ({\rm \bf r}) \ldots
\delta \varepsilon _{rt} ({\rm \bf r}) } \right]E_{t}^{0},
\end{equation}
\begin{eqnarray}
 \overline {{D}_{i}} = {\left[ \varepsilon _{0} \delta _{it} +
\varepsilon _{0} {\sum\limits_{\sigma = 1}^{\infty}\left( { -
{\frac{{1}}{{3\varepsilon _{0}}} }} \right)^{\sigma} \overline
{\mathop {\delta \varepsilon _{ik} ({\rm {\bf r}})\delta
\varepsilon _{km} ({\rm {\bf r}}) \ldots \delta
\varepsilon _{rt} ({\rm {\bf r}})} }}  \right.}\nonumber \\
 + {\left. \sum\limits_{\sigma = 0}^{\infty}  {\left( { -
{\frac{{1}}{{3\varepsilon _{0}}} }} \right)^{\sigma} \overline
{\mathop {\delta \varepsilon _{ik} ({\rm {\bf r}})\delta
\varepsilon _{km} ({\rm {\bf r}}) \ldots \delta \varepsilon _{pr}
({\rm {\bf r}})\delta \varepsilon _{rt} ({\rm {\bf r}})} }}
\right]}E_{t}^{0} . \label{eq:eq14}
\end{eqnarray}
\end{widetext} Note that there are, respectively,  $\sigma $
and $\sigma + 1$ factors under the bars in the sum from 1 to
$\infty $ and that from 0 to $\infty $.

\section{\label{sec:MG} EFFECTIVE PERMITTIVITY OF MATRIX-PARTICLE MIXTURES}

For a macroscopically homogeneous and isotropic mixture, the
averages $\overline {{E}_{i}} $ and $\overline {{D}_{i}} $ are
proportional to the corresponding component $E_{i}^{0} $ of the
external field. Since different terms in series (\ref{eq:eq13})
and (\ref{eq:eq14}) are independent, it is reasonable to suggest
that they have the structure
\begin{equation}
\label{eq:eq15} \overline {\mathop {\delta \varepsilon _{ik} ({\rm
{\bf r}})\delta \varepsilon _{km} ({\rm {\bf r}}) \ldots  \delta
\varepsilon _{rt} ({\rm {\bf r}})} } = A_{\sigma} \delta _{it},
\end{equation}
where the subscript $\sigma$ specifies the number of factors under
the bar. The summation over the indices $i = t$ yields
\begin{equation}
\label{eq:eq16} A_{\sigma}  = {\frac{{1}}{{3}}}\overline
{{\textrm{Tr} }\,\left( {\hat {\mathop {\delta \varepsilon} }
({\rm {\bf r}})\,} \right)^{\sigma }} .
\end{equation}
This trace is easily found, for in the typical expression
\begin{eqnarray*}
\overline {\mathop {\delta \varepsilon _{ik} ({\rm {\bf r}})\delta
\varepsilon _{km} ({\rm {\bf r}}) \ldots
\delta \varepsilon _{rt} ({\rm {\bf r}})} } \\
= {\sum\limits_{\lambda_{1} = 1}^{N} {{\sum\limits_{\lambda_{2} =
1}^{N} {\ldots {\sum\limits_{\lambda_{\sigma} = 1}^{N}
{{\frac{{1}}{{V}}}{\int\limits_{V} {d {\rm {\bf r}}}} \,\Delta
\varepsilon _{ik} ({\rm {\bf r}},{\rm {\Omega}}_{\lambda_{1}} )\Pi
({\rm {\bf r}},{\rm {\Omega}}_{\lambda_{1}}  )}} } }} } \\
  \times \Delta \varepsilon _{km} ({\rm {\bf r}},{\rm
{\Omega}}_{\lambda_{2}}  )\Pi ({\rm {\bf r}},{\rm
{\Omega}}_{\lambda_{2}} ) \ldots \Delta \varepsilon _{rt} ({\rm
{\bf r}},{\rm {\Omega}}_{\lambda_{\sigma}}   )\Pi ({\rm {\bf
r}},{\rm \Omega}_{\lambda_{\sigma}} )
\end{eqnarray*}
all addends with two and more differing values of the indices
$\lambda_{1} ,\lambda_{2} ,\ldots,\lambda_{\sigma} $ are zero (the
regions occupied by hard particles never overlap). If all of these
values are equal, then the relation $[\Pi ({\rm {\bf r}},{\rm
\Omega}_{\lambda} )]^{\sigma}  = \Pi ({\rm {\bf r}},{\rm
\Omega}_{\lambda} )$ ($\sigma $ any natural number) gives
\begin{eqnarray*}
 \overline {\mathop {{\textrm{Tr} }\,\left( {\hat {\mathop {\delta
\varepsilon} }  ({\rm {\bf r}})} \right)^{\sigma }}}  =
{\sum\limits_{\lambda = 1}^{N}
\frac{{1}}{{V}}}{\int\limits_{\Omega_{\lambda}} {d {\rm {\bf r}}}}
\,\Delta \varepsilon _{ik} ({\rm {\bf r}},{\rm \Omega}_{\lambda}
)\Delta \varepsilon _{km} ({\rm {\bf r}},{\rm \Omega}_{\lambda} )
\ldots \\
\times \Delta \varepsilon _{ri} ({\rm {\bf r}},{\rm
\Omega}_{\lambda} )
 = n{\int\limits_{v} {d {\rm {\bf r}}}} \,{\textrm{Tr}}\,\left( {\hat {\mathop
{\Delta \varepsilon} }  ({\rm {\bf r}},{\rm \Omega}_{\lambda} )}
\right)^{\sigma} .
\end{eqnarray*}
\noindent Here, $n = {{N} \mathord{\left/ {\vphantom {{N} {V}}}
\right. \kern-\nulldelimiterspace} {V}}$ is the particle
concentration, the last integral is taken over the region occupied
by a single particle, and the integrands are assumed to be
wise-continuous. Thus,
\[
 A_{\sigma}  = {\frac{{n}}{{3}}}{\int\limits_{v} {d
{\rm {\bf r}}}} \,{\textrm{Tr}}\,\left( {\hat {\mathop {\Delta
\varepsilon} } ({\rm {\bf r}},{\rm \Omega}_{\lambda} )}
\right)^{\sigma},
\]
\noindent or, in terms of the principal components  $\Delta
\varepsilon _{ii} ({\bf{r}})$ of $\hat {\mathop {\Delta
\varepsilon} }$  ($i = 1,\,2,\,3$),
\begin{equation}
\label{eq:eq20} A_{\sigma}  = {\frac{{n}}{{3}}}{\int\limits_{v} {d
{\rm {\bf r}}}} {\{ {[\Delta \varepsilon _{11} ({\rm {\bf
r}})]^{\sigma} + [\Delta \varepsilon _{22} ({\rm {\bf
r}})]^{\sigma}  + [\Delta \varepsilon _{33} ({\rm {\bf
r}})]^{\sigma} \,} \}}, \sigma \ge 1.
\end{equation}

It follows from Eq.~(\ref{eq:eq20}) that the field (\ref{eq:eq13})
and induction (\ref{eq:eq14}) are represented by infinite
geometric series. Summing them up and using definition
(\ref{eq:eq7}), we obtain a formula of the Lorentz--Lorenz type:
\begin{equation}
\label{eq:eq21} {{\varepsilon _{\rm eff}}} = {{\varepsilon _{0}}}
 {\left( {1 + {\frac{{8\pi}} {{3}}}n\alpha _{\rm eff}}\right)} {\left( {1 -
{\frac{{4\pi}} {{3}}}n\alpha _{\rm eff}}\right)}^{-1} {\kern 1pt}
,
\end{equation}
\noindent where
\begin{equation}
\label{eq:eq22} \alpha _{\rm eff} =  {\frac{{1}}{{3}}}\left(
{\alpha _{11} + \alpha _{22} + \alpha _{33}} \right),
\end{equation}
\noindent
\begin{equation}
\label{eq:eq23} \alpha _{ii} = {\frac{{3}}{{4\pi}}
}{\int\limits_{v} {d {\rm {\bf r}}} }{\frac{{\varepsilon _{ii}
({\rm {\bf r}}) - \varepsilon _{0} }}{{2\varepsilon _{0} +
\varepsilon _{ii} ({\rm {\bf r}})}}}, \quad i = 1,\,2,\,3.
\end{equation}

According to Eqs.~(\ref{eq:eq21})--(\ref{eq:eq23}), the effective
polarization properties of an anisotropic particle in a finely
dispersed mixture are described by the quantity $\alpha_{\rm
eff}$. The latter can be treated as the effective polarizability
of the particle in the mixture. The value of $\alpha_{\rm eff}$ is
found as the arithmetic mean (\ref{eq:eq22}) of the quantities
${\alpha _{ii}}$, related to the corresponding principal
components $\varepsilon _{ii} ({\rm {\bf r}})$ of the permittivity
tensor of the particle's substance. However, it is physically
incorrect to interpret ${\alpha _{ii}}$ as the principal
components of a certain tensor, which could be called the
effective polarizability tensor of disperse particles. The
effective polarizability $\alpha_{\rm eff}$ is in fact a scalar
quantity, contributed to by different physical mechanisms; their
combined effect is given by (\ref{eq:eq22}).

 Generalization of Eqs.~(\ref{eq:eq21})--(\ref{eq:eq23}) to the
 case where a mixture
comprises particles of different sorts $a = 1, 2,\ldots, s$, with
individual volumes $v_{a}$ and concentrations $n_{a}$, is evident:
\begin{equation}
\label{eq:eq24} {\varepsilon _{\rm eff}} = {\varepsilon _{0}}
{\left( {{1 + {\frac{{8\pi}} {{3}}}{\sum\limits_{a=1}^{s} {n_{a}
\alpha _{\rm eff}^{a}}} } }\right)}{\left({{1 - {\frac{{4\pi}}
{{3}}}{\sum\limits_{a=1}^{s} {n_{a} \alpha _{\rm eff}^{a}}
}}}\right)}^{-1}{\kern 1pt} ,
\end{equation}
\begin{equation}
\label{eq:eq25} \alpha _{\rm eff}^{a} =  {\frac{{1}}{{3}}}\left(
{\alpha _{11}^{a} + \alpha _{22}^{a} + \alpha _{33}^{a}}  \right),
\end{equation}
\begin{equation}
\label{eq:eq26} \alpha _{ii}^{a} = {\frac{{3}}{{4\pi}}
}{\int\limits_{v_{a}}  {d {\rm {\bf r}}}} {\frac{{\varepsilon
_{ii}^{a} ({\rm {\bf r}}) - \varepsilon _{0} }}{{2\varepsilon _{0}
+ \varepsilon _{ii}^{a} ({\rm {\bf r}})}}}, \quad i = 1,\,2,\,3,
\end{equation}
where $\varepsilon _{ii}^{a} ({\rm {\bf r}})$ are the principal
components of the permittivity tensor of the substance of
particles of sort $a$, and $\alpha _{\rm eff}^{a}$ can be
interpreted as the effective polarizabilities of these particles.

For particles filled with homogeneous anisotropic dielectrics,
Eqs.~(\ref{eq:eq23}) and (\ref{eq:eq26}) take the form (no
summation over $a$)
\begin{equation}
\label{eq:eq27} \alpha _{ii}^{a} = {\frac{{3}}{{4\pi}} }v_{a}
{\frac{{\varepsilon _{ii}^{a} - \varepsilon _{0}}} {{2\varepsilon
_{0} + \varepsilon _{ii}^{a}}} }.
\end{equation}
It is interesting to note that quantities (\ref{eq:eq27}) for an
anisotropic particle are formally equal to the principal
components of the one-particle polarizability tensor for a ball
filled with the same dielectric and having the same individual
volume.

The effective permittivity of a mixture of homogeneous anisotropic
particles is
\begin{eqnarray} \varepsilon _{\rm eff} =
\varepsilon _{0} \left( {1 + {\frac{{2}}{{3}}}{\sum\limits_{a =
1}^{s} {c_{a} {\sum\limits_{i = 1}^{3} {{\frac{{\varepsilon
_{ii}^{a} - \varepsilon _{0}}} {{2\varepsilon _{0} + \varepsilon
_{ii}^{a}}} }}}} }}
\right)\nonumber \\
\,\,\,\, \times \left( {1 - {\frac{{1}}{{3}}}{\sum\limits_{a =
1}^{s} {c_{a} {\sum\limits_{i = 1}^{3} {{\frac{{\varepsilon
_{ii}^{a} - \varepsilon _{0}}} {{2\varepsilon _{0} + \varepsilon
_{ii}^{a}}} }}}} }}  \right)^{ - 1},\label{eq:eq28}
\end{eqnarray}
\noindent where $c_{a}$ is the volume concentration (fraction) of
particles of sort $a$.

For two-component mixtures ($s= 1$), the result (\ref{eq:eq28})
agrees with some of the rules known in the literature. Two
particular examples are of interest. (1) The disperse particles
consist of a substance with isotropic dielectric properties
($\varepsilon _{11}=\varepsilon _{22}=\varepsilon _{33} \equiv
\varepsilon $). Then Eq.~(\ref{eq:eq28}) reduces to the classical
MG mixing rule, no matter what the shapes of the particles are:
\begin{equation}
\label{eq:eq28a} \varepsilon _{\rm eff} =  \varepsilon _{0} + 3 c
\varepsilon_{0}  \frac {\varepsilon - \varepsilon _{0}}{
\varepsilon +   2\varepsilon_{0} -c (\varepsilon -
\varepsilon_{0})}.
\end{equation}
(2) The disperse particles consist of a uniaxial substance, with
one permittivity value ($\varepsilon_{33}\equiv
\varepsilon_{\parallel}$) in one preferred direction and another
in all perpendicular directions ($\varepsilon _{11}=\varepsilon
_{22} \equiv \varepsilon _{\perp}$). Now, Eq.~(\ref{eq:eq28})
takes the form
\begin{eqnarray}
\varepsilon _{\rm eff} =  \varepsilon _{0} + 3 c \varepsilon_{0}
\nonumber \\ \times \frac {(\varepsilon_{\perp}+2
\varepsilon_{0})(\varepsilon_{\parallel} - \varepsilon_{0}) - 2
\varepsilon _{0}(\varepsilon_{\parallel} -
\varepsilon_{\perp})}{(1-c)(\varepsilon_{\perp}+2
\varepsilon_{0})(\varepsilon_{\parallel} + 2\varepsilon_{0}) + c
\varepsilon _{0}(\varepsilon_{\perp} + 2\varepsilon_{\parallel} +
6 \varepsilon_{0})} . \nonumber \\\label{eq:eq28b}
\end{eqnarray}
Eq.~(\ref{eq:eq28b}) first appeared in
Ref.~\onlinecite{bib:Levy1997}, where  an anisotropic version of
the MG approximation was developed for the special case of a
mixture of randomly-oriented uniaxial spherical particles.

It should be emphasized that the mixing rules (\ref{eq:eq21}),
(\ref{eq:eq24}), and (\ref{eq:eq28})--(\ref{eq:eq28b}) were
obtained for sufficiently concentrated mixtures of anisotropic
particles. In deriving them, no constraints on the value of the
difference between the permittivities of the particles and host
were imposed. To evaluate feasible restrictions on concentration
values, we note that these rules were derived for macroscopically
homogeneous and isotropic mixtures with the local permittivity
tensor given by Eq.~(\ref{eq:eq1}). The model (\ref{eq:eq1})
assumes that the particles of a mixture retain their entities to
the greatest extent possible; in other words, they are viewed as
clearly distinct inclusions embedded into a host medium (matrix).
For spherical particles and under certain restrictions (see recent
experimental, \cite{bib:Robinson2002} numerical,
\cite{bib:Mallet2005} and theoretical \cite{bib:Sushko2007}
estimates), the classical MG mixing rule can be accurate for
concentration values as high as $c_{\rm max}^{\rm sph} \approx
(0.4 \div 0.5)$. In the case of nonspherical particles, the shape
effect comes into play: for concentrations exceeding a certain
value $c_{\rm max}$, different orientations of elongated particles
become hindered by the neighboring particles; as a result, the
individualities of such particles start being disturbed. Given the
greatest linear size, $d_{\rm max}$, of a particle, $c_{\rm max}$
can be estimated as $c_{\rm max}  \approx c_{\rm max} ^{\rm sph} v
/ v_{\rm max}$, where $v$ is the volume of the particle and
$v_{\rm max}  = \pi d_{\rm max} ^{3} /6$ is the minimum volume of
an imaginary sphere admitting of free rotations of this particle.
Indeed, if the bulk of each such sphere were completely filled
with the substance of the particles, the above-mentioned results
for mixtures of spherical particle would apply. However, the
actual concentration of the filling substance is $v_{\rm max}/v $
times less than $c_{\rm max} ^{\rm sph}$.

We suggest that the upper bound of the validity range in
concentration for Eqs.~(\ref{eq:eq21}), (\ref{eq:eq24}), and
(\ref{eq:eq28})--(\ref{eq:eq28b}) is no less than $c_{\rm max}$.
For certain packings of particles, even relatively long, and also
for mixtures with wide spreads in the shapes and sizes of
particles, the value of the this bound is expected to increase.

\subsection{Mixtures of ellipsoidal particles}

Consider in detail a mixture of ellipsoids with semiaxes $d_{1} $,
$d_{2} $, $d_{3} $; the ellipsoids are filled with a homogeneous
anisotropic dielectric with permittivity tensor $\varepsilon _{ik}
$. Besides practical applications, this case attracts interest as
an example where the effective polarizability $\alpha_{\rm eff}$
can be expressed explicitly through the geometrical parameters and
the principle components of the one-particle polarizability tensor
of a solitary ellipsoid. These components can be calculated
theoretically and measured experimentally.

For the sake of simplicity, let us suppose that the principal axes
of the tensor $\varepsilon _{ik} $ coincide with the principal
axes of all ellipsoids. The principal components of the
one-particle polarizability tensor of such an ellipsoid in a host
medium of permittivity $\varepsilon _{0}$ are
\cite{bib:Bohren1983,bib:LandauV8}
\begin{equation}
\label{eq:eq30} \gamma _{ii} = {\frac{{1}}{{4\pi}}
}v{\frac{{\varepsilon _{ii} - \varepsilon _{0}}} {{\varepsilon
_{0} + L_{i} (\varepsilon _{ii} - \varepsilon _{0} )}}}, \quad i =
1,2,3,
\end{equation}
\noindent where $\varepsilon _{ii} $ are the principal components
of $\varepsilon _{ik} $ and $L_{i} $ are the depolarization
factors. The latter are given by the integrals
\begin{eqnarray*}
 L_{i} = {\frac{{d_{1} d_{2} d_{3}}}
{{2}}}{\int\limits_{0}^{\infty} {{\frac{{d u}}{{(u + d_{i}^{2}
)R(u)}}}}} , \\
\quad R(u) = \sqrt {(u + d_{1}^{2} )(u + d_{2}^{2}
)(u + d_{3}^{2} )} ,
\end{eqnarray*}
\noindent and satisfy the relation
\begin{equation}
\label{eq:eq32} L_{1} + L_{2} + L_{3} = 1.
\end{equation}

Using Eqs.~(\ref{eq:eq27}), (\ref{eq:eq30}), and (\ref{eq:eq32}),
the desired relation is easy to find:
\begin{equation}
\label{eq:eq33} \alpha _{\rm eff} =
{\frac{{1}}{{3}}}{\sum\limits_{i = 1}^{3} {{\left[ {1 + (1 -
3L_{i} ){\frac{{\gamma _{ii}}} {{d_{1} d_{2} d_{3}}} }} \right]}}}
^{ - 1}\gamma _{ii} .
\end{equation}
If there are different sorts of ellipsoidal particles in a
mixture, Eq.~(\ref{eq:eq33}) and its particular versions
(\ref{eq:eq34}), (\ref{eq:eq36}) hold for each sort separately.

Eq.~(\ref{eq:eq33}) reveals that the effects of two factors of
anisotropy, the shape of the particles and anisotropy of the
dielectric parameters of the particles' substance, on $\alpha
_{\rm eff}$ are interlinked in an intricate way. In the case of a
mixture of anisotropic balls, which is free of the shape
anisotropy effects and where $L_{i} = 1 / 3$, Eq.~(\ref{eq:eq33})
takes the form
\begin{equation}
\label{eq:eq34} \alpha _{\rm eff} = {\frac{{1}}{{3}}}\left(
{\gamma _{11} + \gamma _{22} + \gamma _{33}}  \right).
\end{equation}
\noindent If the ellipsoids are filled with a homogeneous
isotropic dielectric with scalar permittivity $\varepsilon $, then
the principal components of their one-particle polarizability
tensor are
\begin{equation}
\label{eq:eq35} \beta _{ii} = {\frac{{1}}{{4\pi}}
}v{\frac{{\varepsilon - \varepsilon _{0} }}{{\varepsilon _{0} +
L_{i} (\varepsilon - \varepsilon _{0} )}}}, \quad i = 1,2,3,
\end{equation}
\noindent and their effective polarizability in a mixture is
\cite{bib:Sushko2009}
\begin{equation}
\label{eq:eq36} {\alpha _{\rm eff}}  = 3 {\left(
{{\frac{{1}}{{\beta _{11}}} } + {\frac{{1}}{{\beta _{22}}} } +
{\frac{{1}}{{\beta _{33}}} }} \right)}^{-1}.
\end{equation}
\noindent It follows that the shape anisotropy results in a
nonlinear relationship between the effective  polarizability and
the principal components of the one-particle polarizability. For a
mixture of homogeneous isotropic balls, the values of these
polarizabilities are equal.

It should be remembered that results (\ref{eq:eq33}),
(\ref{eq:eq34}), and (\ref{eq:eq36}) are based on the mixing rule
(\ref{eq:eq28}). Except for the case of spherical particles, its
functional form is significantly different from those typical of
one-particle considerations. For instance, one of the most popular
MG-type formulas for a two-component mixture of randomly-oriented
homogeneous ellipsoids reads\cite{bib:Sihvola1999}
\begin{eqnarray}
\label{eq:eq24a}
 \varepsilon _{\rm eff} =  \varepsilon _{0} +
 {\frac {1}{3}} c \varepsilon _{0} {\sum\limits_{i=1}^{3} {\frac {\varepsilon
 - \varepsilon_{0}} {\varepsilon_{0}+ L_{i} (\varepsilon -
 \varepsilon_{0})}}}\nonumber \\
\times \left( 1 - {\frac{1} {3}}c {\sum\limits_{i=1}^{3} {\frac
{L_{i}(\varepsilon -
 \varepsilon_{0})} {\varepsilon_{0}+ L_{i} (\varepsilon -
 \varepsilon_{0})}}} \right)^{-1}{\kern 1pt}.
\end{eqnarray}
A question arises how close are the predictions made by
Eq.~(\ref{eq:eq28a}) [the particular case of Eq.~(\ref{eq:eq28})
for two-component mixtures] and by other mixing rules, such as Eq.
(\ref{eq:eq24a}).

\begin{figure}
\includegraphics[width=7cm]{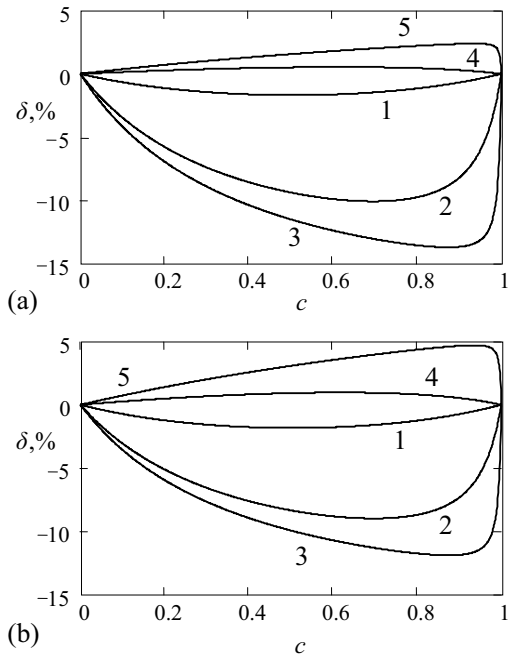}
\caption{\label{fig:f2} $\delta$ as a function of $c$ for mixtures
of (a) prolate spheroids (aspect ratio 2) and (b) oblate spheroids
(aspect ratio 0.5) embedded in a uniform isotropic host. The
permittivities $\varepsilon_{0}$ and $\varepsilon$  are: 2 and 10
(curve 1); 2 and 100 (2); 2 and 1000 (3); 10 and 2 (4); 1000 and 2
(5).}
\end{figure}

To evaluate the numerical discrepancies between
Eqs.~(\ref{eq:eq28a}) and (\ref{eq:eq24a}), we have analyzed the
relative difference $\delta = [(\varepsilon_{\rm
eff}^{(\ref{eq:eq28a})} - \varepsilon_{\rm
eff}^{(\ref{eq:eq24a})})/\varepsilon_{\rm
eff}^{(\ref{eq:eq24a})}]\times 100 \% $ in the values that they
give for $\varepsilon_{\rm eff}$. Fig.~\ref{fig:f2} illustrates
the behavior of $\delta$ for mixtures of prolate ($d_{1}=
d_{2}<d_{3})$ and oblate ($d_{1}= d_{2}>d_{3}$) spheroids with
aspect ratios $d_{3}/d_{1} =2$ ($c_{\rm max}\approx 0.125 $) and
$0.5$ ($c_{\rm max}\approx 0.25 $), respectively.
Several facts are clearly seen for these systems. (1) In the
limiting case of diluted mixtures, $\delta \rightarrow 0$. (2)
Even if the contrast $\kappa \equiv \varepsilon / \varepsilon_{0}$
is extremely high, $|\delta|$ does not exceed $3 \div 4\%$ for $c<
0.5 c_{\rm max}$ and $5 \div 7\%$ for $c \in (0.5 c_{\rm
max},c_{\rm max})$. (3) For moderate values of $c$, $|\delta|$
increases when either $c$, or $\kappa$, or they both increase;
depending on $\kappa$, $|\delta|$ can be as high as $10 \div
15\%$. (4) Compared to Eq.~(\ref {eq:eq24a}), Eq.~(\ref{eq:eq28a})
predicts a smaller magnitude for the ellipsoidal-shape influence
on the formation of $\varepsilon_{\rm eff}$: for a given $c$,
$|\varepsilon_{\rm eff}^{(\ref {eq:eq28a})}-\varepsilon_{0} |<
|\varepsilon_{\rm eff}^{(\ref {eq:eq24a})}-\varepsilon_{0} |$. The
physical explanation for these tendencies is, in our opinion, as
follows: multiple mutual polarizations and short-range
correlations between ellipsoidal particles effectively reduce the
anisotropy effects induced by a single particle. Unfortunately,
the lack of numerical data for the systems under consideration
does not allow us to check directly which of Eqs.~(\ref{eq:eq28a})
and (\ref{eq:eq24a}) does better.

\section{\label{sec:Br}BRUGGEMAN-TYPE MIXING FORMULAS}

As the concentration of anisotropic particles in a mixture becomes
greater than $c_{\rm max}$, two kinds of phenomena can be
expected. On one hand, strong mutual influences can trigger
various physicochemical processes, affecting the textures and
dielectric properties of the particles and host medium. On the
other hand, interparticle contacts can cause the formation of
coarse aggregates of particles; as a result, asymmetry between the
disperse phase and the host medium fades away. In the limit, the
original mixture, including particles of $s$ sorts, can be viewed
as a system of $s + 1$ components, filling chaotically located
regions with irregular complex boundaries.

It was already mentioned that for concentrated mixtures, the
Bruggeman mixing rule is often believed to be superior to the
MG-type mixing rules. However, the accuracy and range of validity
of this rule have remained disputable (for relevant details, see
extensive review \cite{bib:Brosseau2006}). To a great extent, the
situation is explained by the lack of accurate electromagnetic
solutions for disordered systems; as a result, most arguments for
the Bruggeman rule are based on a dipolar approximation, valid for
low concentrations of particles. The purpose of this section is to
shed some light on theses issues. Namely, we show that the
Bruggeman-type formulas are obtainable within the proposed
compact-group approach under appropriate assumptions. In addition,
system-dependent deviations from the classical mixing rules are
expected to occur, provided physical and chemical processes change
the dielectric and structural parameters of the mixture
constituents.

\begin{figure}
\includegraphics[width=5cm]{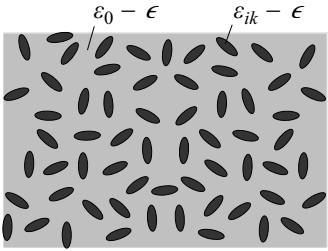}
\caption{\label{fig:f3} Effective medium approximation for the
mixture shown in Fig.~\ref{fig:f1}.}
\end{figure}

To begin with, we remind that Eq.~(\ref{eq:eq13}) and
(\ref{eq:eq14}) are valid for any mixture satisfying the
conditions of macroscopical homogeneity and isotropy. They form
the basis for analysis of particular structural models of
mixtures, which are formulated in terms of the local permittivity
tensor $\varepsilon _{ik} ({\rm {\bf r}})$. For instance, the
above MG-type mixing rules (see Section~\ref{sec:MG}) are based on
the model expression (\ref{eq:eq1}). Now, let us suppose that all
components of a mixture, including the host medium, can be treated
symmetrically in the following sense: there exists a fictitious
(effective) medium, of permittivity $\epsilon$, such that the real
mixture is macroscopically equivalent, in its dielectric
properties, to a macroscopically homogeneous and isotropic system
of nonoverlapping regions occupied by the components of the real
mixture and embedded into this fictitious medium (see
Fig.~\ref{fig:f3}).
 If combining the components into a real
mixture does not change their properties, then the deviations of
the local permittivity in the system from the value $\epsilon$ can
be written as $ \Delta \widetilde{\varepsilon} _{ik}({\rm {\bf
r}},\Omega_{\lambda_{a}})= \varepsilon _{ik} ({\rm {\bf
r}},\Omega_{\lambda_{a}}) - \epsilon \delta_{ik}$ for regions
$\Omega_{\lambda_{a}}$, occupied by particles of sort $a$, and as
$ (\varepsilon _{0}  - \epsilon) \delta_{ik}$ for the rest of
space, occupied by the host. Correspondingly, the local
permittivity deviations $\delta \varepsilon _{ik} ({\rm {\bf r}})$
from $\epsilon$ due to a compact group, of the above regions,
present at the point $\bf {r}$ in the system can be modelled as
\begin{equation}
\label{eq:eq37}  \delta \varepsilon _{ik} ({\rm {\bf r}}) =
(\varepsilon _{0}  - \epsilon) \delta_{ik}\,\Pi ({\rm {\bf r}})+
{\sum\limits_{a = 1}^{s} {\sum\limits_{\lambda_{a} = 1}^{N_{a}}
{\Delta \widetilde{\varepsilon} _{ik} ({\rm {\bf r}},{\rm
\Omega}_{\lambda_{a}} )\,\Pi ({\rm {\bf r}},{\rm \Omega}_{\lambda
_{a}} )}}} ,
\end{equation}
\noindent where \[ \Pi ({\rm {\bf r}}) = 1 - {\sum
\limits_{a=1}^{s} {\sum \limits_{\lambda_{a} = 1}^{N_{a}} \Pi
({\rm {\bf r}},{\rm \Omega}_{\lambda_{a}} )}} \]
 \noindent is the
characteristic function for the space occupied by the host of the
real mixture. Carrying out manipulations similar to those in
Section~\ref{sec:MG} and introducing the total volume
concentration $c = \sum_{a=1}^{s} c_{a}$ of the disperse
particles, we find:
\begin{equation}
\label{eq:eq38} {\frac{{\varepsilon _{\rm eff}}-{\epsilon }}
{2{\epsilon} + {\varepsilon _{\rm eff}} } } = (1-c)
{\frac{{\varepsilon _{0}}-{\epsilon }} {2{\epsilon} + {\varepsilon
_{0}} } } + {\frac{{4\pi}} {{3}}} {\sum \limits_{a = 1}^{s}}
{n_{a}}\widetilde{\alpha} _{\rm eff}^{a},
\end{equation}
\begin{equation}
\label{eq:eq39} \widetilde{\alpha} _{\rm eff}^{a} =
{\frac{{1}}{{3}}}\left( {\widetilde{\alpha} _{11}^{a} +
\widetilde{\alpha} _{22}^{a} + \widetilde{\alpha} _{33}^{a}}
\right),
\end{equation}
\begin{equation}
\label{eq:eq40} \widetilde{\alpha} _{ii}^{a} = {\frac{{3}}{{4\pi}}
}{\int\limits_{v_{a}}  {d {\rm {\bf r}}}} {\frac{{\varepsilon
_{ii}^{a} ({\rm {\bf r}}) - \epsilon }}{{2\epsilon + \varepsilon
_{ii}^{a} ({\rm {\bf r}})}}}, \quad i = 1,\,2,\,3.
\end{equation}
The quantities $\widetilde{\alpha} _{\rm eff}^{a}$ can be
interpreted as the effective polarizabilities of particles of sort
$a$ in the fictitious effective medium of permittivity $\epsilon$.
For mixtures of particles filled with homogeneous anisotropic or
homogeneous isotropic dielectrics, we respectively have:
\begin{equation}
\label{eq:eq41} \ {\frac{{\varepsilon _{\rm eff}}-{\epsilon }}
{2{\epsilon} + {\varepsilon _{\rm eff}} } }=(1-c)
{\frac{{\varepsilon _{0}}-\epsilon} {2\epsilon + \varepsilon _{0}
}} + {\frac{{1}}{{3}}}{\sum\limits_{a = 1}^{s} {c_{a}
{\sum\limits_{i = 1}^{3} {{\frac{{\varepsilon _{ii}^{a} -
\epsilon}} {{2\epsilon + \varepsilon _{ii}^{a}}} }}}} },
\end{equation}
\begin{equation}
\label{eq:eq42}
 {\frac{{\varepsilon _{\rm eff}}-{\epsilon }}
{2{\epsilon} + {\varepsilon _{\rm eff}} } }=(1-c)
{\frac{{\varepsilon _{0}}-\epsilon} {2\epsilon + \varepsilon _{0}
}} + {\sum\limits_{a = 1}^{s} {c_{a} {{\frac{{\varepsilon^{a} -
\epsilon}} {{2\epsilon + \varepsilon^{a}}} }}} }.
\end{equation}

In the above formulas, the permittivity $\epsilon$ is unknown and
should be considered as an adjustable parameter. Under the
additional assumption that the fictitious effective medium is
macroscopically equivalent, in its dielectric properties, to the
mixture itself, $\epsilon =\varepsilon_{\rm eff}$, the left-hand
sides in Eqs.~(\ref{eq:eq38}), (\ref{eq:eq41}), and
(\ref{eq:eq42}) vanish, and we arrive at generalizations of the
classical Bruggeman fixing rule. Note also that particular forms
of Eq.~(\ref{eq:eq42}) are known in the literature (see, for
instance, Ref.~\onlinecite{bib:Musal1988}, where it was obtained
within a dipolar analysis for the case of identical spherical
inclusions). Finally, for $\epsilon = \varepsilon_{0}$, the
results of Section~\ref{sec:MG} are obtained.

\begin{figure}
\includegraphics[width=5cm]{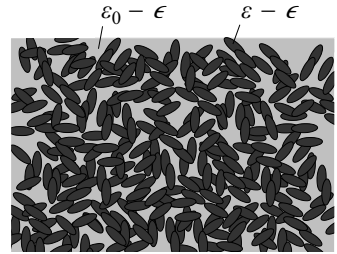}
\caption{\label{fig:f4} Modelling the mixture shown in
Fig.~\ref{fig:f1} for the case of high particle concentrations.}
\end{figure}

For sufficiently concentrated mixtures, large aggregates of
particles can themselves be treated as structural units, with
scalar (on the average) permittivities. Modelling such a mixture
as a system of nonoverlapping regions occupied by substances (the
host medium of the real mixture among them) with different
permittivities and embedded into a fictitious medium of
permittivity $\epsilon$ (see Fig.~\ref{fig:f4}),
 we again arrive
at Eqs.~(\ref{eq:eq38})--(\ref{eq:eq40}), with
$\varepsilon_{ii}^{a}({\rm \bf r}) = \varepsilon^{a}({\rm \bf
r})$, and Eq.~(\ref{eq:eq42}), with $\varepsilon^{a}({\rm \bf r})
=\varepsilon^{a} = {\rm const}$. Now, $\varepsilon^{a}({\rm \bf
r})$ stands for the averaged-permittivity distribution within the
region of sort  $a$, $n_{a}$ is the concentration of such regions,
$v_{a}$ indicates integration over the region of sort  $a$, and
$c_{a}$ stands for the volume concentration of the substance of
sort $a$.

\section{CONCLUSION}

The main results and conclusions of this paper are as follows.

(1) The compact-group method for study of effective dielectric
properties of finely dispersed mixtures is generalized to the case
where the mixtures are composed of anisotropic inhomogeneous
particle. Based on this generalization, new mixing rules for
mixtures of hard dielectric particles are derived and their ranges
of validity are discussed; in appropriate cases, the rules
obtained agree with the classical Maxwell-Garnet and Bruggeman
rules. It should be emphasized that within our method, the
contributions from multiple polarization and short-range
correlation effects of all orders are effectively taken into
account.

(2)  That fact that mixing rules of both types are obtainable
(under appropriate approximations) within a single formalism is
taken by us as strong evidence of the validity of basic relations
(\ref{eq:eq13}) and (\ref{eq:eq14}). If this is really the case,
much emphasis in further effective permittivity studies should be
put on modelling, in terms of the permittivity distribution, of
the structure and dielectric properties of the constituents of a
real mixture. Deviations of these properties from those of
isolated constituents should manifest themselves as deviations
from the classical mixing rules and their modifications obtained
in the present work.

(3) The averages in Eqs.~(\ref{eq:eq13}) and (\ref{eq:eq14}) are
in fact statistical, according to the analysis in
Section~\ref{sec:basic}. Consequently, Eqs.~(\ref{eq:eq13}) and
(\ref{eq:eq14}) can be used to develop a statistical theory of
effective permittivity. In particular, proceeding in this way and
using the general properties of particle distribution functions
for macroscopically homogeneous and isotropic systems of balls,
one can infer that it is the hardness of the balls that lies at
the heart of the classical MG mixing rule.

\begin{acknowledgments}
I thank Prof. V. M. Adamyan for useful discussion.
\end{acknowledgments}

\end{document}